\documentclass[11pt]{article}
\usepackage[margin=1in]{geometry}
\usepackage{amsmath,amssymb,mathtools}
\usepackage{xcolor}
\usepackage{booktabs}
\usepackage{enumitem}
\usepackage{tikz}
\usetikzlibrary{arrows.meta,positioning,shapes.geometric,fit,backgrounds,calc}
\usepackage{listings}
\usepackage[colorlinks=true,linkcolor=blue!55!black,urlcolor=blue!55!black,citecolor=blue!55!black]{hyperref}

\definecolor{codebg}{RGB}{246,247,249}
\definecolor{boxblue}{RGB}{223,235,250}
\definecolor{boxgreen}{RGB}{223,244,228}
\definecolor{boxgray}{RGB}{238,238,240}
\definecolor{boxgold}{RGB}{250,242,214}
\definecolor{edge}{RGB}{120,130,140}

\lstdefinestyle{plain}{
  basicstyle=\ttfamily\small,backgroundcolor=\color{codebg},
  breaklines=true,frame=single,rulecolor=\color{gray!45},
  commentstyle=\color{green!42!black},showstringspaces=false,
  columns=fullflexible,keepspaces=true,xleftmargin=6pt,xrightmargin=2pt}
\newcommand{\ket}[1]{\lvert #1 \rangle}

\newcommand{\braket}[2]{\langle #1 \,\vert\, #2 \rangle}
\newcommand{\Nocc}{N_{\mathrm{occ}}}
\newcommand{\kk}{\mathbf{k}}
\newcommand{\GG}{\mathbf{G}}
\newcommand{\bmb}{\mathrm{BaMg_2Bi_2}}
\newcommand{\bath}{\mathrm{Ba_3BiSb}}

\title{\textbf{Calculation of DFT Spin--Orbit Spillage with Quantum ESPRESSO}}
\author{Duy Quan Nguyen and Paul C.H. Li\\
\small Department of Chemistry, Simon Fraser University, Burnaby, BC, Canada V5A 1S6}
\date{}

\tikzset{
  box/.style={rounded corners=3pt,draw=edge,thick,align=center,
    inner sep=6pt,font=\small,minimum height=1cm,text width=4.2cm},
  bblue/.style={box,fill=boxblue},
  bgreen/.style={box,fill=boxgreen},
  bgray/.style={box,fill=boxgray},
  bgold/.style={box,fill=boxgold},
  flow/.style={-{Stealth[length=3mm]},thick,draw=edge},
}

\begin{document}
\maketitle

\begin{abstract}
\noindent This work describes the calculation of \emph{spin--orbit spillage} from a crystal
structure. Spin--orbit spillage provides a measure of the likelihood that a material has
topological character. The spillage also provides the reference quantity for the machine-learning classifier
of Choudhary et al.~\cite{choudhary2021}, which predicts whether the spillage exceeds a specified
threshold rather than the calculation of its numerical value directly.
The complete computation workflow was applied to the insulating compound $\bmb$, yielding a
spillage of $2.094$ compared with the published VASP spillage of $2.075$, corresponding to a
difference of $0.9\%$. The calculation is described in terms of two Quantum ESPRESSO (QE)
calculations of spillage performed with and without spin--orbit coupling, the role of relativistic
pseudopotentials, and the subsequent wavefunction-overlap analysis. The limitations of the same
calculation procedure for semimetals are also examined.
\end{abstract}

\section{Physical overview}

Band inversion of a material is an important indicator of its topological character~\cite{hasan2010}. Spin--orbit
coupling (SOC), a relativistic effect that is particularly strong in heavy elements such as Bi, can
change the band structure and occupation of electronic states. The \textbf{spillage}~\cite{liu2014}
quantifies the change in the occupied electronic bands upon inclusion of SOC. A small spillage
indicates little change, whereas a large spillage can indicate band inversion and hence a candidate
topological material. Choudhary et al.\ used this quantity to screen thousands of materials in the
JARVIS database~\cite{choudhary2019,choudhary2020npj}.

Evaluation of this change of band structure requires electronic wavefunctions calculated \emph{with}
and \emph{without} SOC. The computational workflow therefore consists of two density-functional-theory
(DFT) calculations followed by a comparison of the corresponding occupied electronic bands.

The reference \texttt{jarvis-tools} implementation reads VASP wavefunction files. Here, the
calculation is implemented directly from QE's per-$\kk$-point HDF5 wavefunction files, providing a
reusable QE workflow and allowing the effects of pseudopotential choice and occupied-band ambiguity
to be examined explicitly.

The representative system is $\bmb$ (JARVIS identifier JVASP-4053), a five-atom hexagonal crystal
($P\bar{3}m1$, $a=b=4.906$~\AA, $c=8.293$~\AA). This compound is an \textbf{insulator}, with a finite band gap
between the highest filled and the lowest empty state at every $\kk$-point, $0.455$~eV with SOC off.
The band gap is narrow once SOC is switched on, $0.040$~eV direct (and $0.021$~eV indirect in
JARVIS), but the gap never closes. As Section~\ref{sec:semi} shows, that single property of non-zero
gap is what makes the calculation
well posed.

\section{Computational workflow}

The diagram below summarizes the computational workflow.

\begin{figure}[htbp]
\centering
\begin{tikzpicture}[node distance=8mm]
  \node[bgray] (struc) {\textbf{1. Crystal structure}\\ $\bmb$, 5 atoms, hexagonal\\ \footnotesize(pulled from the JARVIS database)};

  \node[bblue,below left=10mm and 3mm of struc] (nosoc)
    {\textbf{2a. Non-SOC run}\\ QE SCF, \emph{scalar}-relativistic\\ spinless wavefunctions};
  \node[bgreen,below right=10mm and 3mm of struc] (soc)
    {\textbf{2b. SOC run}\\ QE SCF, \emph{fully} relativistic\\ noncollinear spinors};

  \node[bgray,below=26mm of struc] (wfc)
    {\textbf{3. Wavefunctions (HDF5)}\\ plane-wave coefficients,\\ one file per $\kk$-point, from each run};

  \node[bblue,below=8mm of wfc] (post)
    {\textbf{4. Overlap post-processor}\\ \texttt{compute\_spillage.py}\\ compares the two filled sets};

  \node[bgray,below=8mm of post] (gk)
    {\textbf{5. } $\gamma(\kk)$ \textbf{at every} $\kk$-point\\ \footnotesize how much the filled states changed there};

  \node[bgold,below=8mm of gk] (max)
    {\textbf{6. spillage} $=\displaystyle\max_{\kk}\gamma(\kk)$\\ \footnotesize the one reported number: \textbf{2.094}};

  \draw[flow] (struc) -- (nosoc);
  \draw[flow] (struc) -- (soc);
  \draw[flow] (nosoc) -- (wfc);
  \draw[flow] (soc) -- (wfc);
  \draw[flow] (wfc) -- (post);
  \draw[flow] (post) -- (gk);
  \draw[flow] (gk) -- (max);
\end{tikzpicture}
\caption{Workflow of the spillage computation.}\label{fig:workflow}
\end{figure}
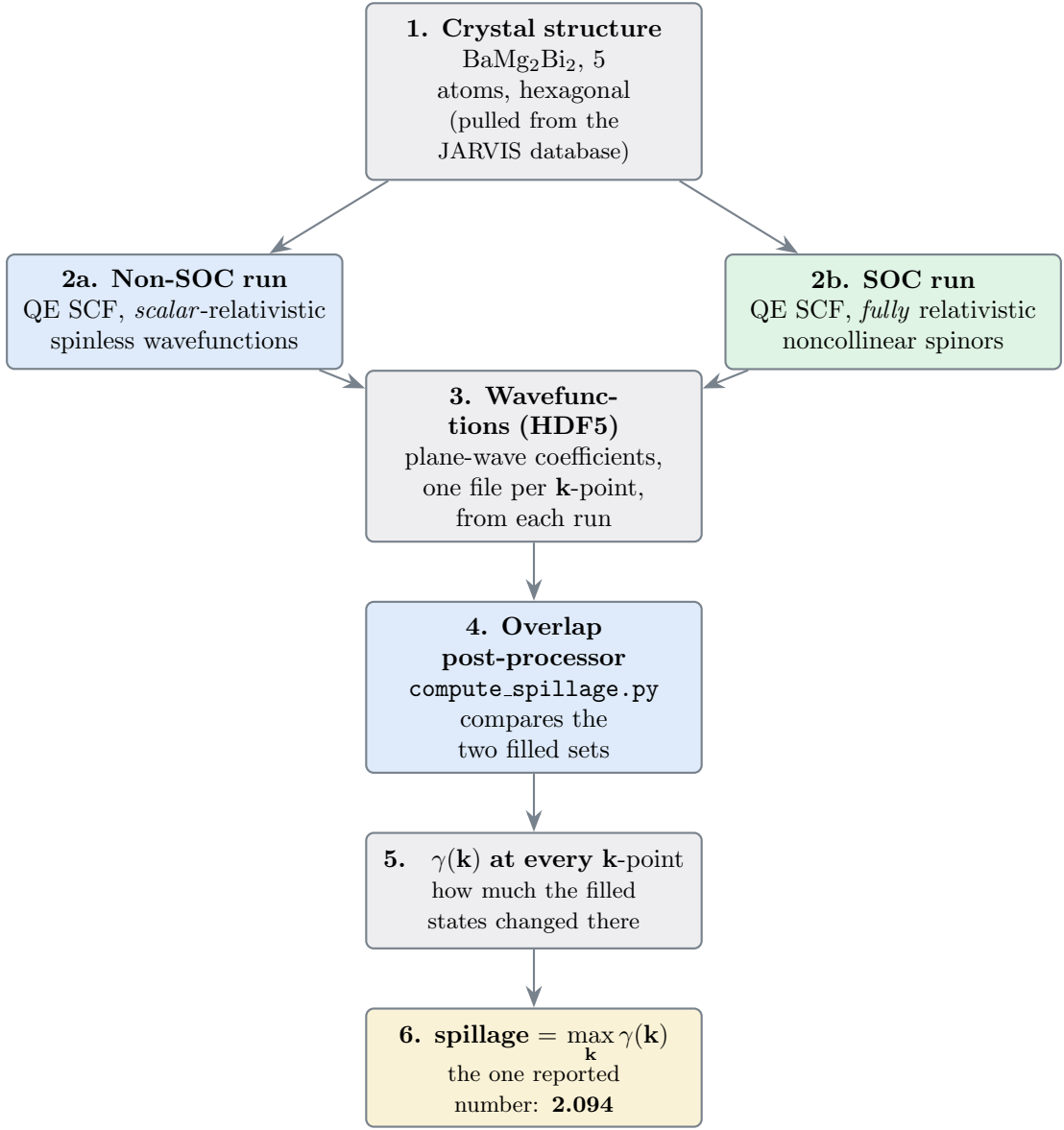

\noindent The two DFT calculations dominate the computational cost. The choice of pseudopotentials
for calculations 2a and 2b requires particular considerations, as discussed in
Sections~\ref{sec:rel} and~\ref{sec:pp}.

\section{Definition of spin--orbit spillage}\label{sec:phys}

In the crystal's momentum space, the ``Brillouin zone,'' DFT generates a list of electronic states
(bands) at each $\kk$-point, arranged in order by energy. The lowest states are occupied and the rest
are empty. Call the set of occupied states without SOC $\{\ket{\psi^{\text{no}}_a(\kk)}\}$ and with
SOC $\{\ket{\psi^{\text{so}}_m(\kk)}\}$. There are $\Nocc$ occupied states in each.

The spillage at $\kk$ is
\begin{equation}
  \gamma(\kk) \;=\; \Nocc \;-\; \sum_{m=1}^{\Nocc}\sum_{a=1}^{\Nocc}
  \bigl|\braket{\psi^{\text{so}}_m(\kk)}{\psi^{\text{no}}_a(\kk)}\bigr|^{2}.
  \label{eq:spillage}
\end{equation}

The double sum measures the overlap between the occupied electronic states obtained by DFT with and
without SOC. If SOC produces no change, every occupied state obtained with SOC aligns with the
occupied state of the non-SOC calculation, the double sum equals the occupied-state count $\Nocc$,
and $\gamma(\kk)=0$. If SOC inverts bands so that occupied and unoccupied bands exchange in energy,
the occupied states no longer coincide, the sum decreases below $\Nocc$, and $\gamma(\kk)>0$.

The material's spillage is the maximum over all $\kk$,
\begin{equation}
  \eta \;=\; \max_{\kk}\,\gamma(\kk),
\end{equation}
because a band inversion may occur at a particular crystal momentum.
Usually the peak is at the $\kk$-point center $\Gamma=(0,0,0)$.

\paragraph{Assumption concerning band occupation.} Equation~\eqref{eq:spillage} assumes a fixed and
unambiguous value of $\Nocc$ at every $\kk$. This condition is satisfied for an insulator but not
necessarily for a metal or semimetal with zero band gap, as discussed in Section~\ref{sec:semi}.

\paragraph{Required calculations.} Equation~\eqref{eq:spillage} requires only the occupied DFT
wavefunctions from calculations with and without SOC evaluated on the same $\kk$-point mesh. These
quantities can be obtained using any plane-wave DFT code, e.g.\ VASP or QE.

\section{DFT calculations with Quantum ESPRESSO (QE)}\label{sec:qe}

QE is an open-source software package~\cite{giannozzi2009,giannozzi2017} that solves the Kohn--Sham
equations of DFT~\cite{hohenberg1964,kohn1965} for a crystalline material. Given a unit cell and
atomic positions, QE computes the DFT electronic energies and wavefunctions of the material. The QE
features relevant to the present calculation are summarized below.

\begin{description}[leftmargin=1.4em,style=nextline]
\item[Plane-wave basis.] The electronic wavefunction of the material is represented by plane waves.
A plane wave is a wave whose value is constant on planes perpendicular to its wave direction.
\item[Pseudopotentials.] Atomic core electrons near the nucleus are represented by spherical waves
and are hard to represent with plane waves. To simplify the wave representation, the nucleus and
core electrons are replaced by an effective potential called a pseudopotential, and only the valence
electrons are treated explicitly. The choice of pseudopotential type is particularly important for
the present calculation, as discussed in Sections~\ref{sec:rel} and~\ref{sec:pp}.
\item[The SCF loop.] The effective potential depends on the electron density being determined. The
self-consistent-field (SCF) loop operates by iterating between solving for the wavefunctions and
updating the density until convergence or self-consistency is achieved.
\item[$\kk$-points.] Material properties are integrals over the Brillouin zone in reciprocal space,
approximated by a finite Monkhorst--Pack~\cite{monkhorst1976} grid of $\kk$-points. A
$6\times6\times4$ grid containing $144$ points and centered at $(0,0,0)$ was used. The mesh is
coarser along $c$ because the cell is longer in that direction and its reciprocal axis is
correspondingly shorter.
\end{description}

\paragraph{Relation between the cutoff and the Fourier representation.} The plane-wave expansion is
a Fourier series. A wavefunction in a crystal is periodic up to the Bloch phase, so
\begin{equation}
  \psi_{n\kk}(\mathbf r)\;=\;e^{i\kk\cdot\mathbf r}\sum_{\GG}c_{n\kk}(\GG)\,e^{i\GG\cdot\mathbf r},
  \label{eq:plane-wave}
\end{equation}
where the complex coefficients $c_{n\kk}(\GG)$ are exactly the Fourier components of the
cell-periodic part and the $\GG$ vectors are reciprocal-lattice vectors. The $\GG$-vector
coefficients are used in the spillage calculations. The wavefunction cutoff truncates the Fourier
series, and \texttt{ecutwfc} defines this truncation. A bigger cutoff means more plane waves and
greater accuracy but higher cost.

Two aspects of the cutoff definition require clarification.

\begin{itemize}[leftmargin=1.4em,itemsep=2pt]
\item \texttt{ecutwfc}~$=70$ means ``keep every plane wave whose kinetic energy
$\hbar^2|\kk+\GG|^2/2m$ is below $70$~Ry.'' In Rydberg atomic units, the condition reduces to
$|\kk+\GG|^2 \le 70$, i.e.\ the $\GG$ vectors inside a
\emph{sphere} of radius $\sqrt{70}=8.37$~bohr$^{-1}$ in reciprocal space. For $\bmb$ that sphere holds
$11{,}573$ plane waves at $\Gamma$. Section~\ref{sec:disk} derives an estimate of this plane-wave
count.
\item \textbf{\texttt{ecutrho} is what sets the fast Fourier transform (FFT).} The SCF loop moves
between real and reciprocal space every iteration, and the loop is implemented with an FFT on a
uniform grid. Reciprocal space is the Fourier transform of real space. The electron density is the
square of the wavefunctions, so its reciprocal components need to extend to twice the wavefunction
cutoff. Because energy is proportional to the square of the reciprocal-vector magnitude, this
requires $\texttt{ecutrho}=4\times\texttt{ecutwfc}=280$~Ry for the norm-conserving default. That
cutoff fixes the FFT box. QE reports a $50\times50\times90$ grid holding $92{,}297$ density
reciprocal-lattice $\GG$-vectors for this cell. The $90$ is along $c$ because the cell is longest
there and therefore needs the most sample points. Section~\ref{sec:disk} derives an estimate of this
count.
\end{itemize}

\section{Calculations with and without spin--orbit coupling}

The spillage requires two SCF calculations of the same crystal structure, one with SOC and one
without SOC. The calculations differ only in the SOC calculation control parameters and the
corresponding pseudopotentials, as summarized in Table~\ref{tab:tworuns}.
The atomic structures were obtained from the JARVIS database after relaxation with OptB88vdW,
and the present SCF calculations use PBE; see Listing~\ref{lst:soc-input}. This matches the procedure of Choudhary
et al.~\cite{choudhary2019}, who evaluated PBE spillage on OptB88vdW-relaxed structures because
OptB88vdW with SOC was not available in VASP.

\begin{table}[htbp]
\centering
\caption{The two SCF calculations. All control parameters are identical except the two SOC switches
and the pseudopotential type required. Calculations 2a and 2b correspond to
Figure~\ref{fig:workflow}.}\label{tab:tworuns}
\begin{tabular}{@{}lll@{}}
\toprule
 & \textbf{Non-SOC run (2a)} & \textbf{SOC run (2b)} \\
\midrule
\texttt{noncolin} & \texttt{.false.} (spin-degenerate) & \texttt{.true.} (spinor wavefunctions) \\
\texttt{lspinorb} & \texttt{.false.} (SOC off) & \texttt{.true.} (spin--orbit term on) \\
pseudopotentials  & scalar-relativistic (\texttt{\_SR}) & fully relativistic (\texttt{\_FR}) \\
complex output & one list & \emph{two} lists (spin up, spin down) \\
bands computed (\texttt{nbnd}) & 48 & 96 \\
\bottomrule
\end{tabular}
\end{table}

\texttt{noncolin=.true.} makes every wavefunction a two-component \emph{spinor} (a spin-up part and a
spin-down part, each its own list of plane-wave coefficients). \texttt{lspinorb=.true.} adds the
actual spin--orbit interaction. Together they are ``the SOC calculation.''

\paragraph{Band counts.} The difference between $48$ and $96$ bands arises from spin degeneracy rather than a
different convergence choice. Without SOC, every spatial orbital
holds one spin-up and one spin-down electron, so $60$ electrons need only $30$ bands and \texttt{nbnd}
$=48$ leaves a $60\%$ margin of $18$ empty bands. With SOC the ordinary spin-up/spin-down occupation
convention is gone: QE considers each one-electron spinor separately, so $60$ electrons need $60$
bands. Kramers pairs can still remain degenerate. Choosing \texttt{nbnd}$=96$ leaves $36$ empty
bands, the same $60\%$ fractional margin. Both runs therefore describe the same $60$ electrons,
although the SOC calculation requires twice as many band labels. This same factor of two reappears
in the post-processor as $\Nocc/2$ versus $\Nocc$ in
Section~\ref{sec:post}.

\paragraph{Consistency between the two calculations.} The cell, the
\texttt{ecutwfc}/\texttt{ecutrho} values, and the $\kk$-point mesh are identical. The settings
\texttt{nosym=.true.} and
\texttt{noinv=.true.} turn off symmetry reduction and time-reversal folding applied
by QE. Symmetry would let QE compute fewer $\kk$-points and reorder plane waves, but SOC lowers the
symmetry differently from the non-SOC run, so the two runs would end up with mismatched $\kk$-point
lists and plane-wave orderings. With symmetry off, both runs use the \emph{same} full $\kk$-grid and
the \emph{same} plane-wave set, so the wavefunctions line up one-to-one and the overlap in
Eq.~\eqref{eq:spillage} is well defined in the two calculations.

\begin{lstlisting}[caption={Relevant parameters in the SOC input file for BaMg2Bi2.},label={lst:soc-input}]
&system
    ecutwfc = 70.0        ! plane-wave cutoff (Ry); MUST match the non-SOC run
    ecutrho = 280.0       ! density cutoff (Ry); MUST match the non-SOC run
    nbnd    = 96          ! how many bands to compute (>= occupied + margin)
    noncolin  = .true.    ! <-- spinor wavefunctions
    lspinorb  = .true.    ! <-- spin-orbit coupling ON  (this is what 2a lacks)
    nosym   = .true.      ! no symmetry reduction, so k-points/G-vectors match run 2a
    noinv   = .true.      ! ... and no time-reversal folding either
    occupations = 'smearing'
    smearing = 'mp'       ! Methfessel-Paxton (mp)
    degauss = 0.01        ! smearing width in Ry (see the caveat on occupations)
/
ATOMIC_SPECIES
  Ba  137.327  Ba_FR.upf  ! <-- "_FR" = fully relativistic. This is where SOC lives.
  Mg   24.305  Mg_FR.upf
  Bi  208.980  Bi_FR.upf
\end{lstlisting}

\paragraph{Fields in \texttt{ATOMIC\_SPECIES} (Ba, Mg, Bi).} The three fields specify the atomic
species label, atomic mass, and pseudopotential filename.

\begin{itemize}[leftmargin=1.4em,itemsep=2pt]
\item The \textbf{atomic label} (\texttt{Ba}) identifies the atomic species in the
\texttt{ATOMIC\_POSITIONS} block. The label is not used for an external chemical lookup. Distinct labels may
therefore be assigned to inequivalent sites of the same element when different pseudopotentials are
required.
\item The \textbf{mass} (\texttt{137.327}) is the isotope-weighted atomic mass in atomic mass units.
It enters calculations involving nuclear inertia or mass weighting, such as molecular dynamics and
vibrational analysis. The mass does not affect the single-point SCF calculations currently performed at fixed atomic
positions.
\item The \textbf{filename} (\texttt{Ba\_FR.upf}) is read as specified from \texttt{pseudo\_dir}. That
file defines the element identity. The file header contains the element symbol, the valence
charge $Z_{\text{val}}$, the exchange--correlation functional for which the atomic species was built, and, for an
\texttt{\_FR} file, the $j$-resolved channels that carry SOC.
\end{itemize}

\noindent For instance, the specified \texttt{.upf} file identifies the element and valence charge
(for example \path{z_valence="15.00"} for Bi), while the numerical tables in the file encode the
ionic potential and valence configuration.

\section{Relativistic treatment}\label{sec:rel}

\subsection{Relativistic effects in electronic-structure calculations}

Electrons bound to a heavy nucleus can move very fast, at an appreciable fraction of the speed of
light, calling for consideration of special relativity. Near a Bi nucleus ($Z=83$), relativistic
effects are therefore significant. The single-electron description formulated by the Schr\"odinger
equation is no longer adequate and should be replaced by the Dirac equation~\cite{dirac1928}, which
represents the electron using a four-component wavefunction.

Expansion of the Dirac equation in powers of $1/c$ yields three corrections to the Schr\"odinger
description. The mass--velocity term ($H_{\mathrm{mv}}$) and Darwin term ($H_{\mathrm D}$) shift the
energies without coupling to spin. The third correction is the \textbf{spin--orbit-coupling term}
($H_{\mathrm{SO}}$),
\begin{equation}
  H_{\text{SO}} \;\propto\; \frac{1}{r}\frac{dV}{dr}\;\mathbf{L}\cdot\mathbf{S},
  \label{eq:soc}
\end{equation}
which couples the spin $\mathbf{S}$ to the orbital angular momentum $\mathbf{L}$ of the electron.
The $H_{\mathrm{SO}}$ term can result in electronic band inversion and topological electronic states.
The $H_{\mathrm{SO}}$ magnitude increases strongly with nuclear charge, making Bi-containing
compounds prominent among candidate topological materials.

\subsection{Scalar- and fully relativistic pseudopotentials}

The distinction between ``scalar-'' and ``fully-'' relativistic treatments differs by keeping the
first two correction terms or all three terms, respectively.

\begin{description}[leftmargin=1.4em,style=nextline]
\item[Scalar-relativistic (SR).] Keeps the mass--velocity and Darwin terms, and \emph{averages
$H_{\text{SO}}$ away}. Specifically, the atomic solve is done in the Koelling--Harmon
scheme~\cite{koelling1977}, which discards the spin--orbit operator and produces states labelled by
electronic orbital angular momentum $\ell$ alone. The result includes relativistic energy corrections but
excludes spin--orbit coupling. Calculation 2a therefore uses \texttt{Ba\_SR.upf}.
\item[Fully-relativistic (FR).] Solves the Dirac equation properly, so states are labelled by the
\emph{total} electronic angular momentum $j = \ell \pm \tfrac12$. A $p$ orbital is no longer one object; it
splits into $p_{1/2}$ and $p_{3/2}$ with different radial potentials, and the size of that splitting
\emph{is} the spin--orbit coupling strength. Storing both $j$ channels is how a \texttt{.upf} file carries
SOC~\cite{bachelet1982,kleinman1980}. Calculation 2b therefore uses \texttt{Ba\_FR.upf}.
\end{description}

When \texttt{lspinorb=.true.}, QE reads those $j$-resolved channels and builds the spin--orbit
operator from the difference between them; the noncollinear/spinor machinery that makes this difference possible
in a plane-wave code is described in Ref.~\cite{dalcorso2005}. With \texttt{lspinorb=.false.} QE
would average the two $j$ channels. An FR pseudopotential and the corresponding SOC flag are both
required.

\subsection{Consequences for the spillage calculation}

\paragraph{Invariance of the plane-wave basis.} The plane-wave set is fixed by the cutoff, the cell, and the
$\kk$-points and is independent of the pseudopotential. So swapping SR for FR leaves the basis
unchanged. The two calculations therefore remain directly comparable, and their difference isolates
the effect of SOC. This unchanged basis is verified numerically in Section~\ref{sec:disk}.

\paragraph{Spinor representation.} Without SOC, each spatial orbital holds one spin-up and one
spin-down electron; QE stores one list of coefficients per band. With SOC, spin and orbital angular
momenta are entangled, no electronic state is purely up or down, and QE stores \emph{two} lists per
band. That doubling is the only structural change the post-processor has to accommodate to compute
the spillage.

\subsection{Post-processing}\label{sec:pyrel}

Relativistic effects are incorporated before post-processing. The PseudoDojo pseudopotentials
used here~\cite{vansetten2018,hamann2013} were generated by solving the Dirac equation for each
isolated atom and storing the result in $j$-resolved channels. QE reads these channels and, when
\texttt{lspinorb=.true.}, assembles the spin--orbit operator
during the SCF cycle. The resulting wavefunctions therefore already contain the relativistic
contributions.

\texttt{compute\_spillage.py} performs linear-algebra operations on the output using
Equations~\eqref{eq:spillage} and~\eqref{eq:spinor-spillage}; see Listing~\ref{lst:spinor-arrays}.

\begin{lstlisting}[caption={Treatment of SOC and non-SOC wavefunction arrays in the post-processor.
\texttt{igwx} is the number of plane waves at this $\kk$-point ($11{,}573$ at $\Gamma$).},label={lst:spinor-arrays}]
# read_wfc(), for EITHER run: the stored array has npol blocks of igwx columns
npol = int(a["npol"])          # 2 for the SOC run, 1 for the non-SOC run
c    = ...                     # shape (nbnd, npol * igwx)

# --- non-SOC run: npol = 1, so there is ONE block and nothing to split ---
Cns = ns["c"][:N_OCC_SPATIAL]                           # (N_occ/2, igwx)

# --- SOC run: npol = 2, so the row is sliced into its two halves ---
up = soc["c"][:, :soc["igwx"]][:, perm]                 # (nbnd, igwx)  spin-up block
dn = soc["c"][:, soc["igwx"]:2 * soc["igwx"]][:, perm]  # (nbnd, igwx)  spin-down block
\end{lstlisting}

\paragraph{Representation of the non-SOC wavefunctions.} The relevant statement is the
\texttt{Cns} assignment. Both calculations
store a band as a row of \texttt{npol}\,$\times$\,\texttt{igwx} complex numbers. With
\texttt{npol}~$=1$, that row is the single spinless coefficient vector and requires no additional
slicing. With \texttt{npol}~$=2$, the row is twice as long and is separated at \texttt{igwx} into
spin-up and spin-down components. Thus, \texttt{npol} determines whether this slicing is required.

\medskip
\noindent Separating the stored array into two components does not itself constitute a relativistic
calculation. Equation~\eqref{eq:soc} is incorporated by the pseudopotential generator, activated in
the QE input, and represented in the wavefunctions read by the post-processing code.

\section{Choice of pseudopotential and overlap metric}\label{sec:pp}

Beyond SR-vs-FR there is a second, independent choice of pseudopotentials, and the choice decides which overlap metric must be
used in Eq.~\eqref{eq:spillage}.

\begin{itemize}[leftmargin=1.4em]
\item With \textbf{norm-conserving (NC)} pseudopotentials, a wavefunction is represented by its
plane-wave coefficients, and $\braket{\psi}{\psi'}$ is the dot product of those two lists.
That dot product is the correct overlap within the norm-conserving pseudo-wavefunction representation.
\item With \textbf{PAW}~\cite{blochl1994} (or ultrasoft) pseudopotentials, the stored wavefunction is a
smoothed version that is missing a piece near each atomic core. The true overlap needs an extra
augmentation correction, which is not contained in the wavefunction file. Omitting this correction
is invalid because even
$\braket{\psi}{\psi}$ is not $1$.
\end{itemize}

A calculation using fully relativistic \textbf{PAW} pseudopotentials was initially conducted on
$\bath$ (see Section~\ref{sec:semi}). The resulting overlap produced an unphysical raw value of
$\gamma\approx-17$, likely caused by a poor augmentation correction. Re-orthonormalization of the
occupied states using a L\"owdin correction~\cite{lowdin1950} approximated the missing augmentation
and yielded $1.63$, but introduced an uncontrolled approximation.

\begin{sloppypar}
This normalization issue was resolved by using \textbf{norm-conserving} FR pseudopotentials (PseudoDojo NC
v0.4~\cite{vansetten2018,hamann2013}, PBE functional~\cite{perdew1996}), so that the ordinary
coefficient dot product is the appropriate overlap and no PAW augmentation correction is needed.
All subsequent results use NC pseudopotentials. For $\bmb$, the valence counts are Ba $10$
($5s^2 5p^6 6s^2$), Mg $10$ ($2s^2 2p^6 3s^2$), and Bi $15$ ($5d^{10}6s^2 6p^3$), giving
$10 + 2(10) + 2(15) = 60$.
\end{sloppypar}

\section{Wavefunction output}\label{sec:disk}

After each SCF calculation, QE writes the wavefunctions to HDF5 files, one per $\kk$-point
(\texttt{wfc1.hdf5}, \texttt{wfc2.hdf5}, \dots). Two wavefunction datasets are used in the overlap analysis.
\begin{itemize}[leftmargin=1.4em,itemsep=1pt]
\item \texttt{MillerIndices}: the list of $\GG$ vectors (which plane waves are present at this $\kk$).
\item \texttt{evc}: the coefficients, one complex number per plane wave per band. In the SOC run each
band has \emph{two} blocks of coefficients (spin up, then spin down), because the wavefunction is a spinor.
\end{itemize}

The common basis of the SR and FR calculations can be verified from these files. At
$\Gamma$, both $\bmb$ plane-wave calculations report exactly $11573$ plane waves. This agreement
follows from the common cutoff rather than the pseudopotential type and permits direct overlap of the
$\GG$-vector coefficients without interpolation.

\paragraph{Estimate of the plane-wave count.} The expected plane-wave count $N_{\mathrm{PW}}$ can be
estimated directly from the cutoff. The count is independent of the $144$ $\kk$-points and $60$
electrons because each $\kk$-point has a basis of approximately this size, and the electron count
determines only the number of occupied bands. The plane-wave count is set by the cell and cutoff.

Counting $\GG$ vectors is counting reciprocal-lattice points inside a sphere. Reciprocal space has one
lattice point per volume $(2\pi)^3/V$, where $V$ is the real-space cell volume, and the cutoff keeps
all reciprocal-lattice points inside a sphere of radius $G_{\max}$ using the wavefunction cutoff
\texttt{ecutwfc}, where $G_{\max}=\sqrt{\texttt{ecutwfc}}$ in bohr$^{-1}$. Thus
\begin{equation}
  N_{\text{PW}} \;\approx\; \frac{\tfrac{4}{3}\pi G_{\max}^{3}}{(2\pi)^{3}/V}
  \;=\; \frac{V\,G_{\max}^{3}}{6\pi^{2}}
  \;=\; \frac{V\,(\texttt{ecutwfc})^{3/2}}{6\pi^{2}}.
  \label{eq:npw}
\end{equation}
QE prints the cell volume of $\bmb$ as $V=1166.38$~bohr$^3$, and
$G_{\max}=\sqrt{70}=8.367$~bohr$^{-1}$, so
\[
  N_{\text{PW}} \;\approx\; \frac{1166.38 \times 585.7}{59.22} \;=\; 11{,}535,
\]
against QE's $11{,}573$, which is $0.3\%$ low. The small shortfall of $N_{\mathrm{PW}}$ is expected.
A sphere cut out of a discrete lattice never contains exactly the continuum estimate, and the
discrepancy shrinks as the sphere grows. Running Equation~\eqref{eq:npw} using the density cutoff
\texttt{ecutrho} is the sharper test, because that sphere is eight times larger:
$\texttt{ecutrho}=280$~Ry gives $G_{\max}=16.733$~bohr$^{-1}$ and predicts $92{,}284$, against the
$92{,}297$ $\GG$-vectors QE reports for the density grid. That is an
error of $0.01\%$.

\section{Wavefunction-overlap post-processing}\label{sec:post}

The script \texttt{compute\_spillage.py} implements Eq.~\eqref{eq:spillage}; see
Listing~\ref{lst:core-spillage}. For each $\kk$-point the code does the following:

\begin{enumerate}[leftmargin=1.6em,itemsep=2pt]
\item \textbf{Read} the non-SOC and SOC wavefunction files for that $\kk$.
\item \textbf{Align} the plane waves by their Miller indices so that coefficient $g$ in one
calculation aligns one-to-one with the same $\GG$ as coefficient $g$ in the other. The correspondence
as controlled by \texttt{nosym} is verified explicitly.
\item \textbf{Take the filled states:} the lowest $\Nocc/2$ spatial orbitals from the non-SOC run and
the lowest $\Nocc$ spinor states from the SOC run. ($\Nocc$ = number of valence electrons, $60$ for
$\bmb$. The non-SOC run is spin-degenerate, so each of its $30$ orbitals stands for a spin-up and a
spin-down state, giving $60$ filled electronic states to match the $60$ SOC ones.)
\item \textbf{Build each non-SOC orbital as a spinor:} $\varphi_a$ becomes $\ket{\varphi_a,\uparrow}$
and $\ket{\varphi_a,\downarrow}$. Then the overlap of a SOC spinor $\ket{\psi^{\text{so}}_m}=
(u_m,d_m)$ with these states is given by $\braket{u_m}{\varphi_a}$ and
$\braket{d_m}{\varphi_a}$.
\item \textbf{Sum} the squared overlaps and subtract from the occupied-state count $\Nocc$ to give
$\gamma$:
\begin{equation}
  \gamma(\kk)=\Nocc-\sum_{m=1}^{\Nocc}\sum_{a=1}^{\Nocc/2}
  \Bigl(\,|\braket{u_m}{\varphi_a}|^2+|\braket{d_m}{\varphi_a}|^2\,\Bigr).
  \label{eq:spinor-spillage}
\end{equation}
\end{enumerate}
These overlaps are plugged into Equation~\eqref{eq:spillage}, resulting in
Equation~\eqref{eq:spinor-spillage}. Finally the script takes $\max_{\kk}\gamma(\kk)$, which is the
material spillage. The script prints two columns, \texttt{gamma\_raw} (dot products) and
\texttt{gamma\_lowdin} (after the re-orthonormalization correction). With NC pseudopotentials the dot
product is the appropriate pseudo-wavefunction overlap. The correction was needed when PAW
pseudopotentials were used.
The two columns must therefore agree apart from numerical error. Their agreement to four decimal
places is used as a regression test of the implementation rather than as evidence for the overlap
metric.

\begin{lstlisting}[caption={Calculation at each $\kk$-point (from
\texttt{compute\_spillage.py}). Every array is (rows = electronic states) $\times$ (columns = plane
waves). For $\bmb$ at $\Gamma$: N\_occ = 60, so the array dimensions are (30, 11573) and
(60, 11573).},label={lst:core-spillage}]
Cns    = nonSOC occupied orbitals      # (N_occ/2, n_planewaves)   <- 30 rows, 11573 columns
up, dn = SOC occupied spinors          # each (N_occ, n_planewaves) <- 60 rows, 11573 columns
A_up   = up.conj() @ Cns.T             # dot product (N_occ, N_occ/2) <u_m | phi_a>
A_dn   = dn.conj() @ Cns.T             # dot product (N_occ, N_occ/2) <d_m | phi_a>
gamma  = N_occ - ( |A_up|^2 + |A_dn|^2 ).sum()
\end{lstlisting}

\paragraph{Electronic-state and plane-wave counts.} The first dimension of each array is the number
of electronic states, whereas the second is the number of plane waves. Only the first dimension is halved. The
plane-wave dimension is $11{,}573$ in both calculations. If the two calculations did not share a
common basis, the matrix product
\texttt{up.conj() @ Cns.T} would not even be conformable.

The halving is a spin-counting statement. The non-SOC run is spin-degenerate, so QE stores each
orbital once and lets it hold two electrons; $60$ electrons therefore occupy $\Nocc/2=30$ orbitals.
The SOC run has no such degeneracy, so QE stores all $60$ occupied spinor states separately. Step 5
in Listing~\ref{lst:core-spillage} reconciles the two: each of the $30$ orbitals is used twice, once
against the spin-up half of a spinor and once against the spin-down half, which is why there are two
overlap matrices \texttt{A\_up} and \texttt{A\_dn} of dimension $(60,30)$ rather than one of
dimension $(60,60)$. Together they carry the same $60\times60$ overlaps required by
Equation~\eqref{eq:spillage} or Equation~\eqref{eq:spinor-spillage}.

\medskip
\noindent Note also that the script fixes $\Nocc=60$ for every $\kk$-point rather than reading QE's
occupation numbers. Section~\ref{sec:occ} justifies that choice for $\bmb$, and
Section~\ref{sec:semi} discusses the limitation of semimetals.

\section{Result for $\bmb$}\label{sec:result}

\begin{table}[htbp]
\centering
\caption{QE results obtained for $\bmb$ (JVASP-4053). The values are compared with published JARVIS
VASP values. The spillage reference is from \texttt{dft\_3d}; the two band-gap references are from
the TopoMoat table.}\label{tab:result}
\begin{tabular}{@{}lccc@{}}
\toprule
quantity & QE (this work) & JARVIS (VASP) & difference \\
\midrule
spillage $\eta=\max_\kk\gamma$ & $\mathbf{2.094}$ & $2.075$ & $+0.9\%$ \\
direct gap, no SOC & $0.455$~eV & $0.451$~eV & $+0.9\%$ \\
direct gap, with SOC & $0.040$~eV & $0.040$~eV & same at shown precision \\
\bottomrule
\end{tabular}
\end{table}

\noindent The calculations used NC pseudopotentials (SR and FR), a $6\times6\times4$ mesh containing
$144$ $\kk$-points, $60$ electrons, and \texttt{ecutwfc} $=70$~Ry. On an Apple M2 Pro processor, the
non-SOC and SOC SCF calculations took $2.3$ and $12.7$ minutes, respectively. The time needed for
post-processing to yield $2.094$ is short.

\paragraph{Analysis of the result.}
\begin{itemize}[leftmargin=1.4em,itemsep=2pt]
\item \textbf{The overlap implementation passes a regression test.} For NC pseudopotentials, the
overlap operator is the identity, $\braket{\phi}{\phi}=1$, and the QE eigenvectors are already
orthonormal under the coefficient dot product. Equality of \texttt{gamma\_raw} and
\texttt{gamma\_lowdin} is therefore expected analytically. Their agreement at all $144$ $\kk$-points
to four decimal places confirms the
plane-wave alignment and array handling, but does not independently validate the overlap metric or
remove the usual finite-cutoff, finite-mesh, and pseudopotential approximations.
\item \textbf{The band structure agrees independently.} The two band-gap results above are not part of the
spillage; they independently verify that the QE electronic structure agrees with the VASP result that
produced the reference. They agree to $1\%$ or better, which gives confidence in the $0.9\%$
spillage agreement. Note also what the
two rows say between them: switching SOC on nearly closes a half-electron-volt gap at $\Gamma$, which
is exactly where $\gamma$ peaks. That is the same reshuffling of the filled states the spillage
quantifies, seen in the band energies instead.
\item \textbf{The sampled peak is sharply localized at $\Gamma$.} The spillage reaches $2.094$ at
$\Gamma$ and falls
to $0.22$ at the nearest sampled $\kk$-points. Because $\Gamma$ remains present on a refined
$\Gamma$-centred mesh, refinement cannot lower this sampled value of $2.094$. It could reveal a larger value at
an intermediate point, however, so formal $\kk$-mesh convergence still requires a denser calculation.
\end{itemize}

\paragraph{Reference value.} JARVIS reports two DFT spillages for JVASP-4053, namely $2.075$ in the
\texttt{dft\_3d} table~\cite{choudhary2019} (the one the ML screen is trained on and the one quoted
above) and $2.081$ in the TopoMoat table~\cite{choudhary2020npj}. The present value of $2.094$ is
$0.9\%$ above the first and $0.6\%$ above the second, so the conclusion does not depend on which is
taken as canonical.

\section{Validation of the occupied-state count}\label{sec:occ}

The post-processing calculation of the spillage of $\bmb$ fixes the occupied-state count
$\Nocc=60$ throughout the Brillouin zone. This choice was validated by direct examination of the DFT
band energies at all $144$ $\kk$-points in both calculations.

\begin{table}[htbp]
\centering
\caption{The band gap above the 60th filled state never closes or becomes zero anywhere in the
Brillouin zone in either run.}\label{tab:gapcheck}
\begin{tabular}{@{}lcc@{}}
\toprule
 & non-SOC ($E_{31}-E_{30}$) & SOC ($E_{61}-E_{60}$) \\
\midrule
smallest gap over all $\kk$ & $0.455$~eV (at $\Gamma$) & $0.040$~eV (at $\Gamma$) \\
$\kk$-points where the gap closes & $0$ of $144$ & $0$ of $144$ \\
\bottomrule
\end{tabular}
\end{table}

\noindent As shown in Table~\ref{tab:gapcheck}, a finite non-zero gap separates the 60th and 61st
states throughout the Brillouin zone in both calculations.
``The lowest 60'' is therefore unambiguous at every $\kk$-point, and $\Nocc=60$ is exact, not an
approximation.

\paragraph{Effect of smearing on the reported occupations.} The input files use Methfessel--Paxton
smearing~\cite{methfessel1989} with width defined by \texttt{degauss}~$=0.01$~Ry~$=0.136$~eV,
inherited from the earlier semimetal calculations. This width exceeds the $0.040$~eV SOC gap, so the
QE occupation numbers at $\Gamma$ come out fractional even though a genuine gap is present, i.e.\
$f=1$ below the gap in the ideal limit. This does \emph{not} affect the reported spillage value
because the post-processor does not use these occupations and instead uses the
fixed count justified above. Table~\ref{tab:occ} reports the QE occupations near the band boundary
and compares them with the values for an ideal insulator.

\begin{table}[htbp]
\centering
\caption{QE occupation numbers $f$ at $\Gamma$ for $\bmb$ near the occupied--unoccupied boundary,
compared with the ideal values of $1$ and $0$ for an insulator. Both calculations report occupations
on a $0$--$1$ scale; in the non-SOC calculation the factor of two for spin is included in the
$\kk$-point weight, not here. Fermi levels: $7.489$~eV (non-SOC), $7.835$~eV
(SOC).}\label{tab:occ}
\setlength{\tabcolsep}{5pt}
\begin{tabular}{@{}llrrcl@{}}
\toprule
calculation & bands & $E$ (eV) & reported $f$ & ideal $f$ & interpretation \\
\midrule
non-SOC & $1$--$27$   & $\le 6.220$ & $1.0000$   & $1$ & ideal occupation \\
        & $28$        & $7.177$     & $1.0027$   & $1$ & MP overshoot \\
        & $29,30$     & $7.238$     & $1.0126$   & $1$ & MP overshoot; $30$ is the last filled \\
\cmidrule(l){2-6}
        & $31$        & $7.693$     & $-0.0279$  & $0$ & MP undershoot; first empty \\
        & $32$--$48$  & $\ge 8.141$ & $-0.0000$  & $0$ & non-ideal occupation \\
\midrule
SOC     & $1$--$56$   & $\le 6.643$ & $1.0000$   & $1$ & ideal occupation \\
        & $57,58$     & $7.729$     & $0.9838$   & $1$ & leakage \\
        & $59,60$     & $7.798$     & $0.7218$   & $1$ & leakage; $60$ is the last filled \\
\cmidrule(l){2-6}
        & $61,62$     & $7.838$     & $0.4826$   & $0$ & first empty; the gap here is $0.040$~eV \\
        & $63,64$     & $8.153$     & $-0.0023$  & $0$ & MP undershoot \\
        & $65$--$96$  & $\ge 9.157$ & $-0.0000$  & $0$ & non-ideal occupation \\
\bottomrule
\end{tabular}
\end{table}

\noindent The table reflects two distinct effects of the occupations.

\begin{itemize}[leftmargin=1.4em,itemsep=2pt]
\item \textbf{The values above $1$ and below $0$ result from the smearing scheme, not the physics.}
Methfessel--Paxton (MP) is not a softened step function. It is a Gaussian function multiplied by
Hermite-polynomial corrections, chosen so that corrected quantities in DFT calculations converge
quickly with smearing width. The price is that the occupation number oscillates slightly outside
$[0,1]$ near the Fermi level. Thus
$1.0126$ and $-0.0279$ in the non-SOC run are the expected signature of MP working as designed. They
are not evidence of a metal: that run has a $0.455$~eV gap, more than three times the smearing width.
A Gaussian or Fermi--Dirac smearing would have stayed inside $[0,1]$ and converged more slowly.
\item \textbf{The $0.72$ and $0.48$ values in the SOC calculation reflect charge leakage.} The gap is $0.040$~eV and
the smearing width is $0.136$~eV, so the smearing genuinely cannot resolve the gap and moves charge
across it. The reported values may therefore be misleading. Interpreted directly, band $61$
appears half-occupied, and a criterion based on $f \ge 0.5$ would count $62$ states as filled rather
than $60$. Section~\ref{sec:semi} shows the resulting discrepancy on materials where the count is
genuinely ambiguous.
\end{itemize}

\noindent These $f$ values are not required to sum to the occupied-state count at an individual
$\kk$-point. At $\Gamma$, the non-SOC values sum to $30.0000$, whereas the SOC values sum to
approximately $60.37$. Only the Brillouin-zone-weighted total is constrained. These $f$ values do not enter the
spillage because $\Nocc$ is fixed at $60$ according to Table~\ref{tab:gapcheck}. Nevertheless, using
\texttt{occupations='fixed'} or a \texttt{degauss} value substantially below the band gap would be more
appropriate for an insulating system.

\section{Limitations for semimetals}\label{sec:semi}

The computational workflow was initially applied using PAW pseudopotentials (PP) to the cubic
semimetal $\bath$ (JVASP-36485) and showed
substantially poorer agreement with the reference value. This discrepancy arises from a limitation
of the spillage definition for semimetals rather than necessarily from the numerical implementation
of the spillage. The discrepancy is larger than for other materials, as shown in
Table~\ref{tab:materials}.

\begin{table}[htbp]
\centering
\caption{Calculated spillage values $\eta$ and comparison with published references.}\label{tab:materials}
\setlength{\tabcolsep}{5pt}
\begin{tabular}{@{}llccccc@{}}
\toprule
material & type & PP type & $\kk$-mesh & $e^-$ & $\eta$ & reference \\
\midrule
$\bmb$ (JVASP-4053) & insulator & SR/FR NC & $6\times6\times4$ & 60 & $\mathbf{2.094}$ & $2.075$\ \ ($+0.9\%$) \\
$\mathrm{Bi_2Se_3}$ (JVASP-1067) & insulator & SR/FR NC & $6\times6\times6$ & 78 & $2.115$ & $2.098$\ \ ($+0.8\%$) \\
$\mathrm{Bi_2Te_3}$ (JVASP-25) & insulator & SR/FR NC & $6\times6\times6$ & 78 & $2.112$ & $2.094$\ \ ($+0.8\%$) \\
$\bath$ (JVASP-36485) & semimetal & SR/FR NC & $4\times4\times4$ & 60 & $2.016$ & $2.267$\ \ ($-11\%$) \\
$\bath$, initial calculation & semimetal & SR/FR PAW & $2\times2\times2$ & 50 & $1.632$ & $2.267$\ \ ($-28\%$) \\
PbTe (JVASP-1103) & insulator & SR/FR NC & $6\times6\times6$ & 30 & $2.010$ & $2.200$~\cite{li2026} \\
\bottomrule
\end{tabular}
\end{table}

The comparison across material classes also contains a $\kk$-mesh-size difference. The $\bath$
semimetal was evaluated on a $4\times4\times4$ mesh, whereas the three reference insulators used
$6\times6\times4$ or $6\times6\times6$ meshes. Because semimetallic states near the Fermi level can
be particularly sensitive to Brillouin-zone sampling, part of the $11\%$ discrepancy may reflect
this difference, and the comparison does not isolate the occupied-band ambiguity alone.

\paragraph{Agreement for clean-gap insulators.} The values calculated for $\bmb$,
$\mathrm{Bi_2Se_3}$, and $\mathrm{Bi_2Te_3}$ exceed the corresponding published VASP values by
$0.9\%$, $0.8\%$, and $0.8\%$, respectively. The consistent sign and magnitude suggest a small
systematic offset between the QE norm-conserving and VASP PAW calculations rather than statistical
variation. An offset below one percent is consistent with the expected agreement between converged
plane-wave codes~\cite{lejaeghere2016}. The $-11\%$ discrepancy for the semimetal is therefore
qualitatively distinct. All three insulating systems also satisfy the condition of
Section~\ref{sec:occ}: the gap above the last filled band never closes anywhere in the zone
($0$ of $144$ $\kk$-points for $\bmb$, $0$ of $216$ for each of the other two).

The $\mathrm{Bi_2X_3}$ calculations ($X=\mathrm{Se}$ or $\mathrm{Te}$) used $216$ $\kk$-points and
occupied-state counts of $78$. Calculations based on fixed counts of $78$ reproduced $2.1146$ and
$2.1116$ to four decimal places, confirming
that the fixed-$\Nocc$ post-processing is insensitive to the occupations reported by QE.

\paragraph{Origin of the discrepancy.} $\bath$ is a semimetal with an indirect gap of
$-0.178$~eV, indicating that the
occupied and empty bands overlap in energy. At $\Gamma$, which is exactly where its spillage peaks,
a \emph{triply degenerate} state lies at the Fermi level. The occupied subspace required by
Eq.~\eqref{eq:spillage} is therefore ill-defined at this point. Shifting the occupation boundary by
two bands produces substantial changes in the calculated value; see Table~\ref{tab:sweepbath}.

\begin{table}[htbp]
\centering
\caption{$\bath$: the same wavefunctions and the same code, with only the assumed occupied count
changed.}\label{tab:sweepbath}
\begin{tabular}{@{}lccccc@{}}
\toprule
electrons assumed occupied & 56 & 58 & \textbf{60} & 62 & 64 \\
\midrule
resulting $\gamma(\Gamma)$ for $\bath$ & $2.12$ & $1.43$ & $\mathbf{2.02}$ & $0.08$ & $1.63$ \\
\bottomrule
\end{tabular}
\end{table}

\noindent For identical wavefunctions and code, changing the assumed occupied-state count by two
bands changes $\gamma$ from $0.08$ to $2.12$. The related compound $\mathrm{Ba_3Bi_2}$
(JVASP-36513) exhibits an even larger sensitivity, as shown in
Table~\ref{tab:sweepba3bi2}.

\begin{table}[htbp]
\centering
\caption{$\mathrm{Ba_3Bi_2}$ (JVASP-36513) against $\bath$ (JVASP-36485), $\gamma(\Gamma)$ as a
function of the assumed occupied-state count. Both are antiperovskite semimetals with a degenerate manifold
on $E_{\mathrm F}$ at $\Gamma$. The values span $0.08$ to about $4.2$; the \texttt{jarvis-tools}
rule selects $62$ electrons and gives $4.12$ for $\mathrm{Ba_3Bi_2}$, against the VASP reference
$4.097$. This near-agreement is included as a diagnostic and is not treated as a validated workflow
result.}\label{tab:sweepba3bi2}
\setlength{\tabcolsep}{6pt}
\begin{tabular}{@{}lcc@{}}
\toprule
electrons assumed occupied & $\bath$ & $\mathrm{Ba_3Bi_2}$ \\
\midrule
$56$                                   & $2.12$ & $2.10$ \\
$58$                                   & $1.43$ & $1.44$ \\
$60$ \ \ (fixed count used here)       & $\mathbf{2.02}$ & $\mathbf{2.78}$ \\
$62$ \ \ (\texttt{jarvis-tools}, occupation $\ge 0.5$) & $0.08$ & $4.12$ \\
$64$                                   & $1.63$ & $4.22$ \\
\midrule
VASP reference                         & $2.267$ & $4.097$ \\
\bottomrule
\end{tabular}
\end{table}

The \texttt{jarvis-tools} criterion counts bands with
occupation $\geq 0.5$ separately at each $\kk$ rather than fixing the count. On
$\mathrm{Ba_3Bi_2}$, the script selects $62$ electrons and yields a spillage of $4.12$ compared with the reference value of
$4.097$, a difference of $0.5\%$. Applied unchanged to $\bath$, the script also selects $62$ electrons but
yields $0.08$ compared with the reference value of $2.267$. The value of $4.12$ is therefore reported
as a diagnostic comparison rather than a validated workflow result. The two materials agree closely at
$56$ and $58$ electrons and only diverge once the count enters the degenerate manifold at
$E_{\mathrm F}$, which is what identifies the manifold as the source of the instability.
Here \texttt{jarvis-tools} refers to the separate JARVIS Python package, not to a component of QE or
the QE conda-forge build.

\paragraph{Implications.} For a clean-gap insulator, the occupied-state count is fixed by the
electronic structure, and the present workflow reproduces the published VASP value within $0.9\%$.
For a zero-gap semimetal, the convention used to define the occupied electronic states can dominate
the result. The
$11\%$ discrepancy for $\bath$ is plausibly dominated by this ambiguity rather than by an incorrectly
inferred general QE--VASP difference. This interpretation is supported by the close agreement generally found between
modern plane-wave codes for ground-state quantities~\cite{lejaeghere2016} and by the $0.9\%$
agreement obtained here for insulators using the same computational procedure. The different
$\kk$-point meshes and material-specific pseudopotential effects remain confounding factors.

\medskip
\noindent This interpretation is not definitive because none of the occupation counts considered above
reproduces the reference value of $2.267$. The closest result is $2.12$ for $56$ electrons. The
analysis establishes that occupation-count ambiguity is sufficiently large to account for the
observed discrepancy, but it does not necessarily establish causality. A definitive assessment would
require mesh-converged calculations and the VASP occupation and degeneracy treatment for the same
structure. What is definite is that both the published VASP result and
the fixed-$60$ QE result support a high-spillage classification for $\bath$, although the numerical
value remains sensitive to the occupied-band convention.

\paragraph{PbTe.} The spillage peaks at the four equivalent
$L$ points, where each value is $2.0103$, and decreases to approximately $0.09$ at neighbouring
points. PbTe has normal band ordering at $L$ and is topologically trivial at ambient pressure, in
contrast to the band-inverted topological crystalline insulator SnTe~\cite{hsieh2012}. The large
spillage therefore indicates a strong SOC-induced change in the occupied electronic states relative to the
scalar-relativistic reference, but it does not establish a band inversion or non-trivial topology.
This distinction illustrates why a high-spillage candidate requires subsequent evaluation of
topological invariants or surface states. Choudhary et al.~\cite{choudhary2019} likewise report rare
high-spillage false positives. Independent confirmation of the numerical value remains necessary.

A recent paper predicts a spillage value of $2.200$ for PbTe~\cite{li2026}, a $9.5\%$ difference
from the present value. Other reported values are $2.320$ for $\bmb$ ($11.8\%$), $2.315$ for
$\mathrm{Bi_2Te_3}$ ($10.6\%$), and $2.306$ for $\bath$ ($1.7\%$)~\cite{li2026}. These differences,
from $1.7\%$ to $11.8\%$, are comparable to those found among the spillage values computed here.

\section{Computational reproducibility}

The input decks, post-processing code, reference calculations, and this document are organized in
the standalone \texttt{qe-spillage/} repository. All commands below are run from the repository
root. The QE input decks are in \texttt{inputs/}, the post-processors are at the repository root,
and the recorded SCF logs and spillage tables are under \texttt{reference/}. QE resolves
\texttt{pseudo\_dir} and \texttt{outdir} relative to the repository root. The untracked
pseudopotentials are placed in \texttt{pseudo\_nc/}, while the two generated \texttt{outdir} paths
are \path{out_nosoc_nc_BaMg2Bi2/} and \path{out_soc_nc_BaMg2Bi2/}, as shown in
Listing~\ref{lst:end-to-end}. The environment
variable \texttt{OMP\_NUM\_THREADS=1} prevents oversubscription when eight MPI processes are used.
The \texttt{-npool 8} option assigns one MPI process to each of eight $\kk$-point pools. These
parallelization settings enhance computational performance but do not affect the calculated
wavefunctions or spillage.

The complete end-to-end command sequence used for the representative calculation is given in
Listing~\ref{lst:end-to-end}.

\begin{lstlisting}[caption={End-to-end commands for the worked example of BaMg2Bi2.},label={lst:end-to-end}]
# 1. Install QE (conda-forge build runs under Rosetta on Apple Silicon)
CONDA_SUBDIR=osx-64 conda create -n qe -c conda-forge qe -y

# 2. Put PseudoDojo NC v0.4 PBE SR/FR files in pseudo_nc/

# 3. From the qe-spillage repository root, run the two SCF calculations
conda activate qe
export OMP_NUM_THREADS=1
mpirun -np 8 pw.x -npool 8 -in inputs/BaMg2Bi2.scf.nosoc.nc.in > nosoc_nc_BaMg2Bi2.out  # ~2 min
mpirun -np 8 pw.x -npool 8 -in inputs/BaMg2Bi2.scf.soc.nc.in   > soc_nc_BaMg2Bi2.out    # ~13 min

# 4. Compute the spillage from the two wavefunction sets
SPILLAGE_NELEC=60 SPILLAGE_REFERENCE=2.075 python compute_spillage.py \
    out_nosoc_nc_BaMg2Bi2/BaMg2Bi2_nosoc_nc.save \
    out_soc_nc_BaMg2Bi2/BaMg2Bi2_soc_nc.save
# -> max gamma = 2.0941 at Gamma
\end{lstlisting}

\begin{sloppypar}
\noindent The files \path{inputs/BaMg2Bi2.scf.nosoc.nc.in} and
\path{inputs/BaMg2Bi2.scf.soc.nc.in} contain the two QE inputs. The root-level script
\texttt{compute\_spillage.py} performs the post-processing. The directory
\texttt{reference/scf-logs/} contains the saved QE outputs, while
\texttt{reference/spillage/} contains the recorded per-$\kk$ spillage tables for all materials.
The \texttt{pseudo\_nc/} and generated \texttt{out\_*/} directories are intentionally excluded from
version control. The LaTeX source and rendered document are in \texttt{docs/}.
\end{sloppypar}

\section{Relation to prior work}\label{sec:prior}

This section distinguishes established methods from the contributions of the present
work.

\paragraph{Established methods.} Spin--orbit spillage was introduced by Liu and
Vanderbilt~\cite{liu2014}, and Equation~\eqref{eq:spillage} corresponds to their equation~4. The use
of spillage as a high-throughput screen for topological materials, including the $\gamma>0.5$ rule
of thumb and the screening of the
JARVIS database, is due to Choudhary and
co-workers~\cite{choudhary2019,choudhary2020npj,choudhary2021}, and the machine-learning classifier
that motivates this work is from Ref.~\cite{choudhary2021}. The $\bmb$ and $\bath$ structures and
their reference spillages are from the JARVIS database~\cite{choudhary2019,choudhary2020}. The
underlying methods are established, and these include Kohn--Sham DFT~\cite{hohenberg1964,kohn1965}, the PBE
functional~\cite{perdew1996}, Quantum ESPRESSO~\cite{giannozzi2009,giannozzi2017}, the Dirac
equation~\cite{dirac1928} and its scalar-relativistic reduction~\cite{koelling1977}, relativistic and
optimized norm-conserving pseudopotentials~\cite{bachelet1982,kleinman1980,hamann2013,vansetten2018},
SOC in a plane-wave pseudopotential code~\cite{dalcorso2005}, PAW~\cite{blochl1994},
Monkhorst--Pack $\kk$-grids~\cite{monkhorst1976}, Methfessel--Paxton smearing~\cite{methfessel1989}, and L\"owdin
orthonormalization~\cite{lowdin1950}.

\paragraph{Brillouin-zone sampling in previous work.} Ref.~\cite{liu2014} specifies
the use of QE, PBE, norm-conserving pseudopotentials from the OPIUM package, and cutoffs of
$55$~Ry for $\mathrm{Bi_2Se_3}$ and $\mathrm{Sb_2Se_3}$ and $65$~Ry for $\mathrm{In_2Se_3}$, ``with an
$8\times8\times8$ Monkhorst--Pack $\kk$ mesh,'' from which the plane-wave wavefunctions are extracted
and their equation~6 is used to evaluate the spillage. Thus, both studies evaluate the DFT
wavefunctions on a regular $\kk$-point mesh, although the computational control parameters differ.

High-symmetry lines are used for visualization rather than Brillouin-zone sampling. Liu and
Vanderbilt used charts and maps to show the momentum dependence of $\gamma(\kk)$. Their Figure~5(a)
traces $\gamma$ along a path through the zone, and Fig.~5(b) shows the spillage as a colour map over
the $(k_x,k_y)$ plane of $\mathrm{Bi_2Se_3}$ at $k_z=0$. The high-symmetry path includes $\Gamma$,
where the inversion is
expected. Section~II~B of Ref.~\cite{liu2014} shows that a topologically non-trivial
system must have $\gamma(\kk)\ge 1$ somewhere in the zone, and for an inversion-symmetric
$\mathbb{Z}_2$ insulator the two partner inversions at $\kk_0$ and $-\kk_0$ merge at a TRIM, so one
expects $\gamma \ge 2$ there.

The high-throughput
screen of Choudhary and co-workers~\cite{choudhary2019,choudhary2021} reduces each material to
$\max_\kk \gamma(\kk)$ for use as a thresholded machine-learning label. The accuracy of our maximum
depends on the sampled $\kk$-point set, motivating the mesh-convergence considerations in
Section~\ref{sec:result}.

\paragraph{Contribution of the present work.} The reference implementation in \texttt{jarvis-tools} and the
high-throughput spillage screens built on it~\cite{choudhary2019,choudhary2020npj,choudhary2021},
obtain wavefunctions from VASP~\cite{kresse1996} by reading \texttt{WAVECAR} files. Computational
reproduction of that workflow therefore requires access to VASP, which is expensive.

QE has previously been used in related stages of the spillage workflow. Liu and Vanderbilt used QE
for their original spillage calculations, as described above. Ref.~\cite{choudhary2019} also used QE
for Wannier interpolation and evaluation of topological invariants after screening 289 candidate
materials, but the spillage itself was calculated with VASP.

No reusable QE-based implementation of the spillage calculation was reported, and no code
accompanies Ref.~\cite{liu2014}. In our present contribution, the post-processor described in
Section~\ref{sec:post} reads QE's per-$\kk$-point HDF5 wavefunction files and enables the calculation
without VASP.

Liu and Vanderbilt report a spillage of $2.12$ at $\Gamma$ for $\mathrm{Bi_2Se_3}$, computed in QE with norm-conserving
pseudopotentials. The present calculation for $\mathrm{Bi_2Se_3}$ (JVASP-1067), using a
$6\times6\times6$ mesh, $78$ electrons, and PseudoDojo NC pseudopotentials, gives $2.1146$ and is also
peaked at $\Gamma$. The difference is $0.3\%$ despite the use of different pseudopotential types,
cutoffs, and meshes, providing direct validation against the work that introduced the quantity.

First, additional results support this agreement. The related compound $\mathrm{Bi_2Te_3}$ (JVASP-25) gives
$2.1116$ compared with its VASP reference of $2.094$. Both compounds peak at $\Gamma$. Their
second-largest sampled values are $0.9851$ and $1.8294$ for $\mathrm{Bi_2Se_3}$ and
$\mathrm{Bi_2Te_3}$, respectively. The latter is only $13\%$ below the sampled maximum, so these
calculations do not establish formal $\kk$-mesh convergence. A refined mesh retains the value at
$\Gamma$ but could identify a larger value at an intermediate point. Repeating both calculations
with \texttt{occupations='fixed'} confirms the values to four decimal places.

The band-gap comparison is less uniform than the spillage comparison. The calculated non-SOC direct
gaps agree with JARVIS, giving $0.279$ versus $0.279$~eV for $\mathrm{Bi_2Se_3}$ and $0.348$ versus
$0.342$~eV for $\mathrm{Bi_2Te_3}$. In contrast, the calculated SOC direct gaps are $0.273$ and
$0.292$~eV for $\mathrm{Bi_2Se_3}$ and $\mathrm{Bi_2Te_3}$, respectively, compared with the JARVIS
values of $0.186$ and $0.223$~eV. The $6\times6\times6$ mesh
likely misses an off-$\Gamma$ SOC gap minimum. This gap discrepancy does not directly alter the
reported spillage at $\Gamma$, which is sampled by both meshes, but it reinforces the need for an
explicit mesh-convergence study.

The $\mathrm{Bi_2Se_3}$ spillage agrees within one percent with both the QE value of $2.12$ reported
by Liu and Vanderbilt~\cite{liu2014} and the JARVIS VASP value of $2.098$~\cite{choudhary2019}.
Together with the systematic offset
discussed in Section~\ref{sec:semi}, these results indicate that the QE implementation reproduces
published spillage values for systems with well-defined occupied subspaces, subject to the stated
mesh limitation.

The pseudopotential choice in Section~\ref{sec:pp} has a methodological consequence. With PAW pseudopotentials
the true overlap requires the augmentation operator $S$, which is not stored in the wavefunction
file. After DFT, the \texttt{jarvis-tools} implementation handles this augmentation by
orthogonalizing the occupied coefficient matrices numerically via SVD, under the assumption that the
SOC and non-SOC wavefunctions span the same space. This orthogonalization is an uncontrolled
approximation and changed the FR-PAW result from
a raw $\gamma\approx-17$ to $1.63$. Choosing \emph{norm-conserving} pseudopotentials removes the need
for PAW augmentation in this overlap: the coefficient dot product in
Equation~\eqref{eq:spinor-spillage} is then the appropriate pseudo-wavefunction overlap.
Consequently, agreement of \texttt{gamma\_raw} and
\texttt{gamma\_lowdin} is expected for NC wavefunctions and does not validate the metric
independently. Their agreement at every $\kk$-point is retained as a regression test for
plane-wave alignment and basis consistency. Liu and Vanderbilt also used norm-conserving
pseudopotentials, so their Eq.~(6) likewise uses the ordinary coefficient overlap.

\begin{sloppypar}
The underlying principle is established. Klime\v{s}, Kaltak, and
Kresse~\cite{klimes2014} show, in the context of $GW$ quasiparticle energies, that PAW describes
overlap integrals incorrectly because the partial-wave basis inside the atomic spheres is incomplete,
and that the error ``can be avoided by adopting norm-conserving partial waves.'' The same mechanism
applies here to the spillage overlap. The present work shows that the pseudopotential type determines
the appropriate overlap metric in Eq.~\eqref{eq:spillage}. The
\texttt{gamma\_raw}/\texttt{gamma\_lowdin} comparison provides a regression test of the
implementation rather than independent evidence for that metric. To the best of our knowledge, the
connection between pseudopotential type and the spillage overlap metric has not been stated explicitly
in the spillage literature.
\end{sloppypar}

The numerical results obtained in the present work are $\eta=2.094$ for $\bmb$ compared with the
$2.075$ reference, $2.016$ and $1.632$ for $\bath$ using NC and PAW pseudopotentials, respectively,
and $2.010$ for PbTe, compared with a reported value of $2.200$~\cite{li2026}. The analysis in
Section~\ref{sec:semi} demonstrates that, for a semimetal with a degenerate manifold at the Fermi
level, the spillage is ill-conditioned with respect to the occupied-band count and varies between
$0.08$ and $4.2$ across plausible choices. This sensitivity is a
general limitation of the method rather than one specific to the present implementation.

The two $\bath$ runs also changed the cutoff, $\kk$-mesh and valence-electron count, so the difference
between the corrected values $1.632$ and $2.016$ cannot be assigned to the pseudopotential alone. The
raw PAW result near $-17$ nevertheless establishes that an ordinary coefficient dot product does not
provide the correct PAW overlap in the absence of the augmentation metric.

\section*{Author contributions and AI-assisted editing}

Duy Quan Nguyen carried out the calculations, implemented the reproducibility workflow, analyzed the
results, and prepared the initial manuscript. Paul C.~H.~Li supervised the project, contributed to
the interpretation of the results, and reviewed and edited the manuscript. Both authors approved the
final manuscript.

AI-assisted tools were used solely for language and \LaTeX{} editing. All scientific content,
calculations, analysis, and conclusions were produced and verified by the authors.

\section*{Glossary}
\addcontentsline{toc}{section}{Glossary}
\begin{description}[leftmargin=0em,style=nextline,itemsep=2pt]
\item[DFT / Kohn--Sham (K--S)] Density functional theory (DFT)~\cite{kohn1965}, a widely used method
for calculating electronic properties (wavefunctions and density) of solids. DFT maps the interacting
many-electron problem onto a set of effective single-electron Kohn--Sham (K--S) equations.
\item[SCF] The self-consistent-field (SCF) loop is the iterate-until-stable loop QE uses to solve the
K--S equations in DFT to self-consistency or convergence.
\item[Spin--orbit coupling (SOC)] A relativistic effect coupling the spin to the orbital angular
momentum, a magnetic effect, of an electron, Equation~\eqref{eq:soc}. Strong in heavy atoms (Bi, Pb).
SOC is what can invert bands and create
topological materials.
\item[Dirac equation] The relativistic wave equation for an electron~\cite{dirac1928}. Spin--orbit
coupling drops out of it as a low-velocity correction; it is solved per atom when a fully-relativistic
pseudopotential is generated.
\item[Scalar- vs fully-relativistic (SR / FR)] SR keeps the relativistic energy corrections but
averages spin--orbit coupling away; FR keeps it, storing separate $j=\ell\pm\frac12$ channels.
The two calculations use \texttt{\_SR.upf} and \texttt{\_FR.upf} files, respectively. See
Section~\ref{sec:rel}.
\item[Spinor] A wavefunction with a spin-up and a spin-down component. SOC calculations use spinors.
\item[Band inversion] When SOC swaps the character of an occupied and an empty band. It is an
important indicator of possible topological character, not by itself a proof, and is what large
spillage is designed to detect.
\item[Spillage $\gamma(\kk)$, $\eta$] Equations~\eqref{eq:spillage} and
\eqref{eq:spinor-spillage}: how much the filled electronic states change when SOC is turned on.
$\eta=\max_\kk\gamma$ is the reported per-material spillage value.
\item[Pseudopotential] An effective potential replacing the nucleus and core electrons, so only
valence electrons are computed. NC and PAW are two types of pseudopotentials (PP); SR and FR PP are
without and with SOC, respectively.
\item[Plane-wave cutoff (\texttt{ecutwfc}, \texttt{ecutrho})] Energy limits setting how many plane
waves represent the electron wavefunctions (wfc) and density (rho). A higher cutoff is more accurate
but more costly. Note these
are energies in Rydberg, not counts: \texttt{ecutwfc}~$=70$ keeps every $\GG$ with
$|\kk+\GG|^2 \le 70$, which for $\bmb$ has a count of $11{,}573$ plane waves.
\item[$\GG$ vectors] The reciprocal-lattice vectors of the crystal, and the frequencies of the
Fourier series in which every wavefunction is expanded. Which ones are present is decided by the
cutoff (a sphere in reciprocal space) and listed in the wavefunction file as
\texttt{MillerIndices}, the integer triples $(h,k,l)$ that label each one. The identical cell, cutoff
and matched $\kk$-points give the same plane-wave basis in the two runs; the identical crystal
structure and \texttt{nosym}/\texttt{noinv} ensure that
QE writes the same full $\kk$-point list rather than reducing it differently. See Eq.~\eqref{eq:npw}
for counting the $\GG$-vectors.
\item[Plane-wave coefficients] The list of complex numbers $c_{n\kk}(\GG)$ multiplying each plane
wave; see Equation~\eqref{eq:plane-wave}.
A single band at a single $\kk$-point \emph{is} this list, stored in the wavefunction file as
\texttt{evc}. With norm-conserving
pseudopotentials, the overlap $\braket{\psi}{\psi'}$ of Eq.~\eqref{eq:spillage} is the dot
product of two such lists. In the SOC run each band carries two of these lists back to back, one per
spin component.
\item[FFT grid] The uniform real-space and reciprocal-space grids that QE transforms between during
each SCF iteration. QE distinguishes smooth and dense grids. The density cutoff \texttt{ecutrho}
controls the dense grid,
reported here as $50\times50\times90$. The cutoffs are therefore parameters of the Fourier
representation.
\item[Monkhorst--Pack mesh] The regular grid of $\kk$-points used to sample the Brillouin
zone~\cite{monkhorst1976}. The mesh used here is $6\times6\times4$, giving $144$ points. This grid is
distinct from a high-symmetry \emph{path}, which is a line through selected points used for plotting
band structures and, in Ref.~\cite{liu2014}, for plotting $\gamma(\kk)$.
\item[Occupation numbers $f$] The per-band fillings QE writes under
\texttt{occupation numbers} in its output, on a $0$--$1$ scale in both calculations. For an ideal
insulator, $f$ would be exactly $1$ below the gap and $0$ above it. With Methfessel--Paxton (MP)
smearing it is not: the scheme can overshoot above $1$ and go slightly negative by construction. If
the smearing width exceeds the gap, the charge is also transferred across the gap. See
Table~\ref{tab:occ}. The
post-processor does not read these values and instead uses a fixed $\Nocc$ justified by the band energies.
\item[MPI / $\kk$-point pool] MPI (Message Passing Interface) is the protocol that lets several
processes work on one DFT calculation. A pool is a group of processes assigned its own subset of
$\kk$-points (\texttt{-npool}). This pool arrangement can work well when there are many $\kk$-points,
but the best pool count must be timed. Pool parallelism
does not distribute the main real- and reciprocal-space arrays, so increasing the number of pools can
increase memory use. The pool changes the speed and memory layout, never the physical answer.
\item[SVD (singular value decomposition)] SVD is a factorisation of a matrix into two unitary
matrices $(V,W)$ and a non-negative diagonal matrix $\Sigma$ of singular values. Liu and Vanderbilt
apply SVD in their valence-band-resolved spillage analysis to the overlap matrix
$L_{nm}(\kk)=\braket{\psi_{n\kk}}{\tilde\psi_{m\kk}}$ connecting occupied states \emph{without} SOC to
\emph{unoccupied} states \emph{with} SOC. Writing $L=V\Sigma W^{\dagger}$ and transforming the two
sets by $V$ and $W$ makes the overlap matrix between
them real and diagonal, which is the gauge in which the band-resolved spillage is read off. Note this
is a different matrix from the occupied-occupied overlap $M_{mn}(\kk)=\braket{\psi_{m\kk}}{\tilde\psi_{n\kk}}$
of their equation~5~\cite{liu2014}, which is the one Eq.~\eqref{eq:spillage} above is built from and to which no SVD is
applied. Separately, \texttt{jarvis-tools} uses SVD to orthogonalise PAW coefficient matrices
numerically, also as a stand-in for the augmentation operator $S$. That second use is an
approximation. With norm-conserving pseudopotentials it is unnecessary, and the related
L\"owdin transformation changes no values at the reported precision of four decimal places.
\item[$\kk$-point / Brillouin zone] The Brillouin zone is the crystal's momentum space, which is the
reciprocal space of real space; $\kk$-points are the sample grid over reciprocal space.
$\Gamma=(0,0,0)$ is its center.
\item[TRIM points] The time-reversal-invariant momenta (TRIM): high-symmetry $\kk$-points ($\Gamma$, and
zone-boundary points) where band inversions often occur. For an inversion-symmetric insulator, the
Fu--Kane parity criterion evaluates the $\mathbb{Z}_2$ indices from parity eigenvalues at the
TRIM~\cite{fu2007}; more general systems require other formulations.
\item[Semimetal] A material whose occupied and empty bands overlap in energy, so there is no gap or
zero band gap. The spillage is not uniquely defined; see Section~\ref{sec:semi}.
\item[Smearing width (\texttt{degauss})] A small numerical softening of the filled/empty boundary,
needed for metals so the SCF is stable. For an insulator the width should be narrower than the band
gap. Ours is
Methfessel--Paxton~\cite{methfessel1989} at $0.01$~Ry~$=0.136$~eV, which is narrower than the non-SOC
gap but wider than the SOC one, hence Table~\ref{tab:occ}.
\item[\texttt{nosym} / \texttt{noinv}] Turn off symmetry reduction and time-reversal folding, so the
two SCF runs share an identical basis of $\kk$-points and plane waves. The SOC and non-SOC
wavefunctions can then be overlapped and compared to compute the spillage.
\item[L\"owdin orthonormalization] A standard way to force a set of states to be mutually orthonormal~\cite{lowdin1950}.
The orthonormalization was used to approximate the missing PAW augmentation. With NC
pseudopotentials this is unnecessary.
\end{description}

\end{document}